# A Quantum-safe Key Hierarchy and Dynamic Security Association for LTE/SAE in 5G Scenario

Rajakumar Arul, *Member, IEEE*, Gunasekaran Raja, Alaa Omran Almagrabi, Mohammed Saeed Alkatheiri, Chauhdary Sajjad Hussain and Ali Kashif Bashir, *Senior Member, IEEE*

*Abstract*— Millions of devices are going to participate in 5G producing a huge space for security threats. The 5G specification goals require rigid and robust security protocol against such threats. Quantum cryptography is a recently emerged term in which we test the robustness of security protocols against Quantum computers. Therefore, in this paper, we propose a security protocol called Quantum Key GRID for Authentication and Key Agreement (QKG-AKA) scheme for the dynamic security association. This scheme is efficiently deployed in Long Term Evolution (LTE) architecture without any significant modifications in the underlying base system. The proposed QKG-AKA mechanism is analyzed for robustness and proven safe against quantum computers. The simulation results and performance analysis show drastic improvement regarding security and key management over existing schemes.

## I. INTRODUCTION

Wireless communication has now become an essential component in our day-to-day life. It is expected to rise by a factor of 500-1000 times in next ten years. According to the recent researches, the intensity of the wireless has been confirmed, which shows that the broadband wireless networks has been adopted in the improvement of education, health care, Industry establishment, and society enrichment [1-3]. The wireless broadband services that support high-speed data transfer helps to establish reliable communication through voice and video over the internet. This communication makes our life so simple and comfortable. Long Time Evolution (LTE) Technology, the primary driver of the 5G wireless network developed by 3rd Generation Partnership Project (3GPP) is a rapidly expanding global standard for the choice for 5G deployments around the universe [4]. LTE is mainly planned to offer extended bandwidth - up to 10x for mobile devices, latency reduction, low maintenance and mobility improvement. LTE-A and evolving technologies like IoT with new enrichment are the vital driving force of 5G technology that offers essential features to support performance and cost optimization in IoT application. IoT with the integration of cellular communication will ensure extraordinary development in next few years for the investors and device manufacturers. According to GSMA research and prediction, mobile IoT communications are expected to hit for over 10% of the global market by 2020. But as enormous devices are connected on wireless network and IoT, concerns are abuzz about security issues associated with these networks. These developments offer new opportunities, but the same time brings advanced challenging security risks for the integrity environment [5].

Service providers are facing more complex issues as they tend to grow and "future-proof" their networks to handle traffic, scaling capability, orchestration, cost control, security, etc. Though LTE is considered to be secured, the architecture has uncovered risks in security verticals. The security issues will bring the overall subscribers experience down and financial losses. Preserving the "key" used by the cryptographic algorithm is the challenging task that persists in the telecom industry [6-9]. Hence, security and key management are the critical issues for the 3GPP activity on LTE in IoT environment. In LTE network if the source key is compromised, the whole set of keys and network become vulnerable is accompanied by the disadvantage of substantial dependency on single key [11], Therefore, in LTE network, providing a safe communication and maintaining the Quality of Service (QoS) are the primary concerns. In this aspect, Quantum cryptography offers an absolute solution towards secured communication over the network by encoding messages as polarized photons, which can be transferred through the air. Quantum Computing combines two of the main scientific achievements of the 20th century: Information Theory and Quantum Mechanics. Its interdisciplinary character is one of the most stimulating and appealing attributes [13, 14].

Nowadays, communications and transactions seem to be safe as they are completely insulated by encrypting with complicated cryptosystems. Quantum Computing makes this belief false, as many of the complex crypto suits like RSA, Elliptic curve cryptography are demonstrated to be vulnerable

Rajakumar Arul is with Department of Computer Science and Engineering, Amrita School of Engineering, Bengaluru, Amrita Vishwa Vidyapeetham, India (e-mail: rajakumararul@ieee.org)

Gunasekaran Raja is with the NGNLab, Department of Computer Technology, Anna University, Chennai 600025, India (e-mail: dr.r.gunasekaran@ieee.org).

Alaa Omran Almagrabi is with Department of Information Systems, Faculty of Computing and Information Technology (FCIT) King Abdul Aziz University (KAU) Jeddah, Kingdom of Saudi Arabia (e-mail: aalmagrabi3@kau.edu.sa)

Mohammed Saeed Alkatheiri and Chauhdary Sajjad Hussain is with College of Computer Science and Engineering, University of Jeddah, Jeddah 21589, Saudi Arabia (e-mail: msalkatheri@uj.edu.sa, shussain1@uj.edu.sa)

Ali Kashif Bashir is with Department of Computing and Mathematics, Manchester Metropolitan University, Manchester, M15 6BH, United Kingdom.(e-mail: dr.alikashif.b@ieee.org)

to quantum computers [15], [16]. Any crypto suite that was considered safe based on discrete log problem and integer factorization will be considered vulnerable against attacks launched using quantum computers. Not only the future encrypted data are unsafe, but even the information captured over the radio and stored, may also be highly vulnerable to a quantum computer. It is highly recommended for the industry to deploy quantum-safe algorithms so that their confidential information can be maintained safe. Thus, we propose a simple key generation and key management scheme, which takes the benefits of Quantum Key Distribution (QKD) as a seed for the key generation and provides a quantum-safe key hierarchy in the 5G scenario for interconnected IoT devices.

The rest of the paper is organized as, section 2 deals with the survey of LTE and Quantum cryptography. In section 3, we present the QK-GRID generation using QKD. A key management solution for quantum-safe key hierarchy mechanisms and its components are given shortly in section 4. The mathematical model and the performance analysis of the proposed methods are conferred in section 5 and section 6 respectively. Finally, the conclusion of the present work is discussed in section 7.

## II. RELATED WORKS

The LTE network is the promising technology of this decade for seamless connectivity. Security is a serious concern in the evolving wireless networks. The security architecture of LTE comprises of key generation and key management through a specific framework called Evolved Packet System Authentication and Key Agreement (EPS-AKA). Though EPS-AKA is good in terms of handling various security loopholes, a permanent security association still seems to be a major security flaw i.e. Single permanent key is used to derive all the future keys [17, 18]. Degefa [19] proposed new approach without any additional cost in EPS-AKA environment wherein, instead of fetching the authentication vectors from home network, fetching authentication vector from the foreign network is enabled to improve the overall performance. This method reduces the message overhead and authentication delay significantly.

In LTE communication, the handover states involving a Home eNodeB (HeNB) is complicated, and key chaining used for this process is considered to lack in security. Hence, in order to manage the handover efficiently in HeNB, Yue [20] proposed a proxy signature-based handover scheme, which is based on Elliptic Curve Cryptography algorithm. This method reduces the computational cost compared with other handover schemes. Hassanein et al., [23] proposed a new authentication and key agreement protocol based on EAP-AK. This protocol combines Elliptic Curve Diffie-Hellman with symmetric key cryptosystem to overwhelm the different types of attacks in EAP-AKA. This method provided perfect forward secrecy to improve the authentication between User Equipment (UE) and AAA server and between the UE/Home Subscriber Server (HSS). Zhao et al. [22] discussed several security issues of the LTE network and concluded that there are still many security threats exists in LTE environment[27].

David Deutsch [24] proposed a Quantum Computing model which used the states of light to solve the complex problems by combining quantum physics and quantum mechanics. Later quantum computer is built and is formulated for very specific purposes. By the evolution of the quantum computers, many problems that are believed to be hard are solved in polynomial time and which created an awareness in the cryptography domain. The predominant number theory problems that pave the way for cryptography is discrete logarithm and integer factoring. These problems are widely believed to be the hard problem, and because of their hardness, these form the base of many cryptographic algorithms. Most of the asymmetric cryptographic algorithms and symmetric key exchange mechanisms, rely on these problems and their security level is determined based on the hardness it provides. That is, time and space the algorithm requires to solve this problem. Shor [13] devised a Quantum Computing algorithm, which solves this discrete logarithm and integer factorization problems in an unexpected time frame reduced by $2/3^{rd}$ of the classical computer. Any conventional algorithm that fits either deterministic or probabilistic will take an average of n/2 matches to find a searched element which is further reduced to $1/4^{th}$ in the quantum computer [25]

Grover LK [26], proposed a quantum algorithm by adjusting the phase of various operations on superposition states that provide the result in less than $\log(\sqrt{N})$. Cryptosuits likes RSA and ECC are proven to be not quantum-safe as they cannot be employed by increasing their key size which out spaces the development of the quantum computer [24]. For example, to attack a 3072-bit RSA key, a quantum computer may require more than 1000 logical qubits. Though it's practically difficult to build such computer, once it's constructed, RSA becomes bait to it. Though doubling the key size will make it safe, the running time of the algorithm is increased by the factor of 8 in a classical computer. The cryptographic algorithms that are safe from attack posted by quantum computers are referred as Quantum-Safe algorithms. Quantum computer made a revolution in cryptography and made "Quantum-Safe" algorithm, a mandate to have secure communication [28]. In this context, we propose a Quantum-Safe key hierarchy and a fruitful solution (QKG-AKA) to solve the permanent single key association, which makes the LTE system safe even from the quantum computers. The proposed model combines the limitations of the quantum computer and provides a robust solution with minimal upgradation in the existing architecture of the LTE networks.

*QKD in LTE environment*

As LTE is deployed globally, these services create new opportunities and revenues but the same time brings new security threats like injection, eavesdropping, content modification and exhaustive resource utilization. Existing research works in the LTE security management schemes focus on the multiple key management techniques, minimization of key distribution and cost reduction mechanisms. Yet, these benefits are realized for the scenario of the static key only, which unfortunately is accompanied by the disadvantage of heavy dependency on single key. To overcome this issue, we come up with an efficient dynamic key generation mechanism that uses the limitations of the quantum computer such as i) Knowledge of the quantum state is always ambiguous, ii) No-cloning principle, iii) No-deletion principle, and iv) No-flipping principle to increases the security as well as utilizes the keys efficiently. The proposed QK-GRID formed as result of QKD process utilizes the properties of quantum mechanics and provides a Quantum-safe key hierarchy for the LTE Network from which the derived keys are generated and processed.

### III. PROPOSED LTE ARCHITECTURE WITH QK-GRID

In order to prevent the confidential information during transmission, a security framework is required for end-to-end communication to ensure the LTE based IoT networks to stay secured. Even though LTE have multiple strong and vigorous cryptographic algorithms [25] still there is a need for improvement, especially in the Key Hierarchy. Subsequently, many research and efforts have been made to find new reliable cryptographic methods. One of these findings has led to the improvement of quantum cryptography, whose security trusts not on expectations about computer power, but on the laws of quantum physics [21, 30]. Though many quantum cryptographic methods have been offered, a perfect secrecy in key distribution mechanism named Quantum Key Distribution-using quantum physics phenomena is well experimented and analyzed [28].

As shown in Fig. 1, the proposed work in LTE/SAE environment comprises four main components: E-UTRAN, UE, EPC and USIM/AuC. The E-UTRAN is responsible for entire radio management in LTE [25]. It consists of eNBs and UE, where eNB is responsible for the resource management to UE that comprise the IP header and encrypt the data stream. The EPC consists of Mobility Management Entity (MME), Serving GateWay (SGW), Packet Data Network Gateway (PDN GW), HSS and Policy and Charging Rules Function (PCRF). The MME is the control entity for all the control plane operations. SGW is used for routing and forwarding the packets also it performs as a local mobility entity for inter eNB handovers. PGW takes charge of all the IP packet based operations like packet inspection, IP allocation, etc. HSS is a global database which contains all mobile users and subscriber related data's and PCRF is for policy and charging control decision-making. In USIM/AuC, the authentication secret key "K" is stored in the USIM and AuC. The USIM creates entity authentication between the network and the user. In this component, our novel QK-GRID is generated by using the quantum key distribution channel and is used for future key generation. The generation of QK-GRID novel Quantum Key Distribution system is illustrated in the following section.

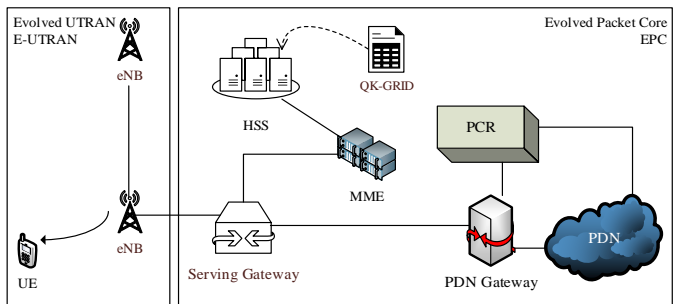

Fig. 1. LTE Architecture with QK-GRID

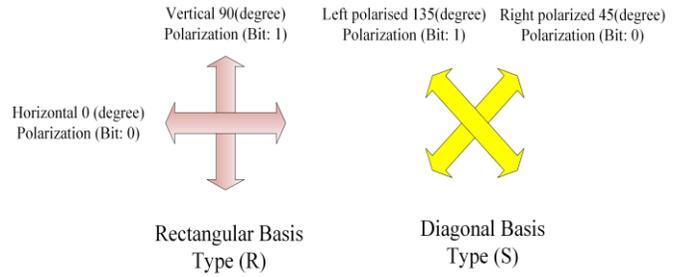

Fig. 2. Basis Types used in QK-GRID Generation

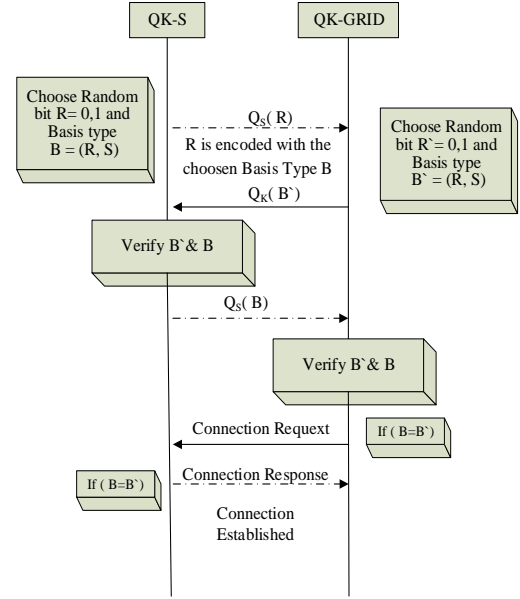

Fig. 3. Qbits generation using QKD

### 3.1 Quantum Key GRID (QK-GRID) generation using QKD

QKD mechanism is for secured shared key generation between two entities. A new quantum key GRID generation scheme based on QKD for safe key hierarchy and a dynamic security association in LTE/SAE environment has been proposed. This mechanism is defined by two different entities such as Quantum Key-Server (QK-S) and Quantum Key GRID (QK-GRID) for single or entangled quantum transmission. The QK-S is a secret key generation, management and exchange server using QKD protocols that are connected by a quantum channel to QK-GRID. The QK-GRID acts as a seed for key generation from the LTE key hierarchy. If the QK-S wishes to establish a secret key with QK-GRID, it begins with the principle of quantum physics for unconditional security [16]. First, the QK-S sends a set of random bits to QK-GRID which are selected from the two states namely Basis Type (R) and Basis Type (S) that uses the photon polarization based on rectangular and diagonal as shown in Fig. 2. Light contains particles called photons which brings electromagnetic energy and intrinsic angular momentum that directs its polarization properties. The polarization of the light is taken by the way photons spin. In the formation of QK-GRID, we considered two basis type named R and S. In R Basis Type, we state a binary 0 as a polarization of 0 degree and 1 as 90 degree. Likewise, a binary 1 can be 135 degree and 0 as 45 degree polarization in the S Basis Type. A single photon will be polarized with four different positions: horizontal h, vertical v, left side polarized lp, and right side polarized rp. Hence, bits can be generated from any one of the two-basis type signified by polarizing the photon.

The following phases shown in fig. 3 explains the process of QKD in the QK-GRID construction. In the first phase, QK-S and QK-GRID setup a quantum channel for key distribution and in the second phase, using the classical channel shared secret key will be recovered as discussed below.

### A. Phase 1 - Quantum communication

In the first phase, QK-S generates a random bit (0 or 1) and chooses a Basis Type R or S, for encoding the bit. QK-S sends a polarized photon for each bit over the Quantum channel to QK-GRID. Upon receiving the photon, the QK-GRID will randomly choose a Basis Type and generates a random bit. After receiving the qubits, QK-GRID communicates its Basis Type of selection to QK-S. QK-S responds back, finding in which Basis Type agreed with QK-GRID's basis. If they use different Basis Type, the measurement results are eliminated and only if the measurement is correct, it is considered as perfect transmission. While Phase I focuses on Basis Type generation, in Phase II QK-GRID will broadcast the basis of measurement.

### B. Phase II - Public discussion

In the second phase, the QK-GRID will report to QK-S about the choice of basis type for each bit and the QK-GRID will notify back whether it made the same choice or not over a traditional insecure public channel. In this point, if a QK-GRID measure different photon, the QK-S and QK-GRID will reject the bits belonging to the particular photons with 50% probability. If no errors occurred or no manipulations happened in the photon, QK-S and QK-GRID confirm that both have an identical string of bits which is known as a sifted key. Information reconciliation is used to calculate the error rate and disclose the occurrence of snooping. The error rate can be calculated in many sources like environmental noise, non-idealities in equipment and eavesdroppers, etc., which can mismatch the keys of QK-S and QK-GRID. Eavesdroppers can collect only partial information through the quantum channel and on the public channel during the information reconciliation process. The knowledge of eavesdroppers can be reduced by the use of universal hash functions that have randomly chosen a value from a public set of functions [18]. As given in Fig. 3, the QK-GRID selects the right basis type, then it will measure the exact photon. If it picks the wrong basis type, then the result will be mismatched. Hence, the ideal efficiency of this selection mechanism is 50%.

### 3.2 Formal operation of Quantum Key –GRID (QK-GRID) generation.

The formal procedure of the QKG-AKA mechanism for layout preparation to create the QK-GRID is represented in Algorithm 1. The conditions for the layout construction are as follows, i) outline of the grid is a $nxn$ matrix and n should be odd always, ii) randomly φ value should be placed on the constructed layout, iii) each row and column should not have more than one φ value and iv) each column has at most one mirror and is varied by the bit size it holds.

| Algorithm 1: QK- GRID Layout Construction |
|---|
| **Input:** Variable bit layout, Null values |
| **Output:** QK-GRID as a table |
| 1: **Start** *QK-GRID Layout Construction* |
| 2: **Initiate** *'nxn' matrix as GRID Layout // n should be odd* |
| 3: Setup GRID as variable bit columns |
| 4: **for** 'i'th column |
|         (2^3*i) sets the Storage capacity of the column |
| 5:   **for** each column **do** |
| 6:     null (φ) value placement |
| 7:   **endfor** |
| 8:   **for** each row **do** |
| 9:     **if** (position != NULL) |
| 10:       set position = QK_Key(column_Size) |
| 11:     **endif** |
| 12:   **endfor** |
| 13: **endfor** |
| 14: **Stop** GRID formation |
| 15: **Stop** QK-GRID Layout Construction |

| Algorithm 2: QK- GRID Generation |
|---|
| **Input:** Quantum key, QK-GRID Layout |
| **Output:** QK-GRID with binaries |
| 1: **Start** *QK-GRID Generation* |
| 2: **Start** Quantum exchange phase |
| 3: **Initiate** *QK-S* |
| 4: Select [rand(0,1), rand(R,S)] |
| 5: Send [rand(0,1), rand(R,S)], rand(photon_sequence) to QK- GRID |
| 6: **In QK-GRID, select rand(R,S)** |
| 7: If (QK-S_rand(R,S) == QK-GRID_rand(R,S)) |
| 8:     **Then** establish_channel **end if** |
| 9: **Stop** Quantum exchange phase |
| 10: **Start** Public discussion phase |
| 11: QK-S, Send (rand(0,1)) for selected *R, S* to QK-GRID |
| 12: QK-GRID, Send (rand(0,1)) for selected *R, S* to QK-S |
| 13: **If** (QK-S_rand (0,1) == QK-GRID_rand(0,1)) |
| 14:     **Then** start connection **end if** |
| 15: **Stop** Public discussion phase |
| 16: **Start** Binaries_Fetch |
| 17:   Send Binaries to QK-GRID |
| 18:   Receive Binaries from QK-S |
| 19: **Stop** Binaries_Fetch |
| 20: Flood Binaries |
| 21: **Stop** *QK-GRID Generation* |

After the layout construction, the values inside the QK-GRID are filled up using QKD process. The generation of QK-Grid values through all the three phases, Quantum exchange, Public discussion and shared secret key are represented in Algorithm 2.

Initially, the QK-S and QK-GRID establish a connection between them by randomly choosing the Basis Type and bits.

*3.2.1 Eavesdropper detection*

For instance, if an eavesdropper is trying to intercept the transmissions from QK-S, it randomly picks qubits based on measurement and retransmits it. 50% of the time adversary would be correct since he does not aware of the Basis Type what QK-S transmitted in and 50% of the time QK-S and QK-GRID are correct when they are retransmitting and measuring the bits. During the sifted key process, the combination of these will present 25% error. Hence, if QK-S and QK-GRID detect an error rate of 25% or higher in sifted key, they conclude that there is an eavesdropper present and weaken the key exchange process. By this way QK-S and QK-GRID always detect the existence of an eavesdropper. Thus, QKG-AKA provides a protective shield against various attacks.

## IV. KEY MANAGEMENT SOLUTIONS

Typically, data encryption is used for data confidentiality, in order to comply with industry regulations. There are multiple options generally used for generating an encryption key, from open source and vendor provided solutions. Once an encryption key is generated and leveraged for encrypting data, the symmetric key has to be stored for decryption. The prime issue that seeks attention is the place where we store the generated key and the environment where it is stored. One of the popular approaches is to use a key hierarchy [6] that simplifies the way to organize encryption keys. Also, the key hierarchy gives a pattern for storing cryptographic keys. It further allows you to derive varying keys that are required for maintaining confidentiality and integrity. Point to be noted is that a master key which is used as a seed for all the derivatives has the power to decrypt all other keys, thereby indirectly all other data. So, protection of master key becomes important; but it is also important that the master key is accessible for decryption. It is highly complex to balance both the availability and protection of master key.

To overcome such a static and single key issue, here a key hierarchy mechanism is proposed. We took a permanent PSA as Previous Key (Pre-K) and a Quantum based key generation component QK-GRID. The QK-GRID is formed of bits that are obtained through a specially designed Quantum based key distributor. This QK-GRID act as a source key material to fetch keys that are needed to make the primary Key LTE K a dynamic and a promising key for all communications.

As shown in Fig. 4, the new key hierarchy of security keys used and the relationship among these security keys are summarized as follows:

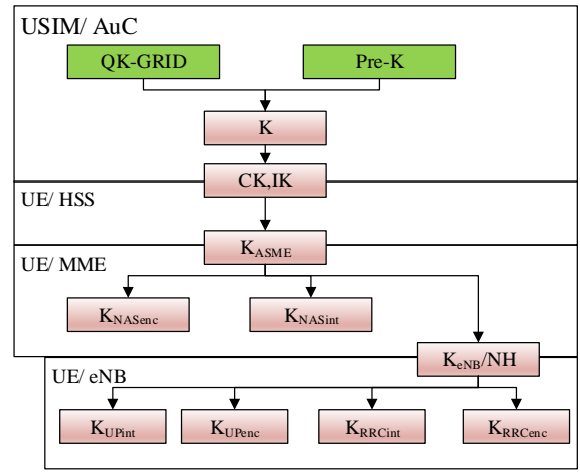

Fig. 4 Proposed LTE Key hierarchy with QK-GRID

- QK-GRID is a table of binaries obtained as a result of QKD process. The main focus of QK-GRID is for primary key establishment and authentication dynamically. Pre-K is the previously used 'K' for all the key generations.
- The master secret key "K" for mobile communication (GSM/EPS) which is stored permanently in USIM and AuC (Authentication Centre) node. "K" is a pre-shared secret for Authentication and Key Agreement (AKA) between the AuC and USIM.
- CK, IK – a pair of keys for confidentiality and integrity.

They are derived and passed from USIM to UE and AuC to HSS during AKC process.
- $K_{ASME}$ – is an intermediate key generated by UE and HSS using keys CK and IK.
- $K_{NASenc}$, $K_{NASint}$ – Encryption and integrity protection generated from $K_{ASME}$ derived in UE/MME.
- $K_{eNB}$ – Intermediate key generated by MME/UE from $K_{ASME}$.
- NH is an intermediate key derived by UE/eNB which provides forward security.

The authentication and key management scheme using QK-GRID is named as QKG-AKA. The QK-GRID is a table that holds variable sized bits across the rows and columns. These bits are initially formed as the result of QKD process. When a user is joining a network, this QK-GRID is formed and is set up based on steps mentioned in Algorithm 1.

QK-GRID is a randomly generated table for users, and need not be unique for each user. The primary key for the LTE is derived from the QK-GRID as shown in Fig. 5. This key is fetched from two components - (1) QK-GRID and (2) Pre-K feedback from past used key K. It is pre-installed for the first run, with a random 256 bits, and for subsequent runs it will use previously used K and in future it will be placed as Pre-K, thereby holding a two-stage security, since key is fetched from two different sources as presented in the Fig. 5. In addition, many manipulation operations are done before obtaining the next key K. The generated LTE master key should be agreed by all the authenticated subscribers and further keys can be generated from this LTE master key to protect users from various interfaces and network. Thus, by adopting QKG-AKA, the static single key permanent association is avoided and dynamic key generation is supported by using QK-GRID. Also, QKG-AKA promises forward and backward secrecy because of QKD and using a feedback circuit for Pre- Key. With the integration of QKD process, QKG-AKA promises a secure key management and GRID formation [6] delivers a quantum-safe key derivation seed.

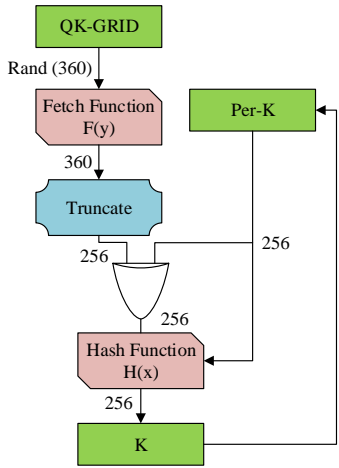

Fig. 5 Generating LTE key through QGK Process

## V. MATHEMATICAL ANALYSIS

In this section, we propose an analytic model to investigate the lifetime of the key as it is the seed for all the derivatives. The parameters used in this analytical model are presented in the Table 1. Let σ, τ and $\lambda_R$ be the expiration rate of the key in a particular QK-GRID session, the mean value of its expiration time and the refreshing rate in the dynamic session respectively. The key expiration time $t_e$ is independent and identically distributed random variables with Cumulative Distribution Function (CDF) as $F_e(t_e)$ and Probability Distribution Function (PDF) as $f_e(t_e)$.

For this, the laplace transform will be

$$f_e^*(s) = \int_{t=0}^{\infty} e^{-st} f_e(t) dt$$

TABLE I
LIST OF PARAMETERS

| Description | Parameter |
|---|---|
| The expected key arrival rate | λ |
| Key expiration time | $t_e$ |
| The expected value of $t_e$ | T |
| PDF of $t_e$ | $f_e(t_e)$ |
| CDF of $t_e$ | $F_e(t_e)$ |
| Laplace transform of $t_e$ | $f_e^*(s)$ |
| Refreshing rate | $\lambda_R$ |
| Time interval between refreshing operation | $t_R$ |
| Re-use probability | P |
| Key discontinuation rate | φ |

The time interval between the key enters into the expiration period and the time when the first refreshing operation occurs will be denoted as $t_r$. Then the probability that atleast single static key generation happens in the expiration period of the key is p. Therefore

$$p = P(t_e > t_r) = 1 - P(t_e \leq t_r) \quad (1)$$

The probability that no key generation happens within the expiration period $t_e$ is given as

$$P(t_e \leq t_r) = \int_{t=0}^{\infty} P\{t_e \leq t_r | t_r = t\} f_r(t) dt$$

$$= \int_{t=0}^{\infty} P\{t_e \leq t | t_r = t\} f_r(t) dt$$

$$= \int_{t=0}^{\infty} P\{t_e \leq t\} f_r(t) dt$$

$$= \int_{t=0}^{\infty} F_e(t) f_r(t) dt$$

$$P(t_e \leq t_r) = \int_{t=0}^{\infty} \int_{t_e=0}^{t} f_e(t_e) f_r(t) dt_e dt \quad (2)$$

Using Excess Life Theorem in [29], [30] $f_r(t)$ is the PDF and can be written as

$$f_r(t) = \lambda_R \int_{s=t}^{\infty} f_R(s) ds$$

$$= \lambda_R [1 - F_R(t)] \quad (3)$$

In the dynamic key generation algorithm, the refreshing rate $\lambda_R$ can be expressed as

$$\lambda_R = \sqrt{\frac{\lambda}{2\varphi_{th}}}$$

The PDF of this refreshing operation is given by

$$p(m) = \frac{1}{T_s}\sqrt{\frac{2\varphi_{th}}{\lambda}}, m \in N^0, m \in \left[0, \left\lfloor T_s \frac{\lambda}{2\varphi_{th}} \right\rfloor\right]$$

The CDF is given as

$$F_R(t) = \begin{cases} \frac{m}{T_s}\sqrt{\frac{2\varphi_{th}}{\lambda}}, m\sqrt{\frac{2\varphi_{th}}{\lambda}} \leq t < (m+1)\sqrt{\frac{2\varphi_{th}}{\lambda}} \\ 1, (m+1)\sqrt{\frac{2\varphi_{th}}{\lambda}} \leq t \leq T_S \end{cases} \quad (4)$$

Combining (3) and (4), (2) is written as

$$P\{t_e \leq t_r\} = \int_{t_r=0}^{\sqrt{\frac{2\varphi_{th}}{\lambda}}} \int_{t_e=0}^{t_r} f_e(t_e) f_r(t_r) dt_e dt_r$$

Thus (refresh time interval $\geq t_r$)

$$= \int_{t_r=0}^{\sqrt{\frac{2\varphi_{th}}{\lambda}}} \int_{t_e=0}^{t_r} f_e(t_e) \lambda_R [1 - F_R(t_r)] dt_e dt_r$$

$$= \int_{t_e=0}^{\sqrt{\frac{2\varphi_{th}}{\lambda}}} \int_{t_r=t_e}^{\sqrt{\frac{2\varphi_{th}}{\lambda}}} f_e(t_e) \lambda_R [1 - F_R(t_r)] dt_r dt_e$$

$$= \int_{t_e=0}^{\sqrt{\frac{2\varphi_{th}}{\lambda}}} \int_{t_r=t_e}^{\sqrt{\frac{2\varphi_{th}}{\lambda}}} f_e(t_e) \lambda_R dt_r dt_e$$

By (4),

$$F_r(t_r) = 0, \text{ if } 0 \le t_r \le \sqrt{\frac{2\varphi_{th}}{\lambda}}$$

$$= \int_{t_e=0}^{\sqrt{\frac{2\varphi_{th}}{\lambda}}} \lambda_R f_e(t_e) \left(\sqrt{\frac{2\varphi_{th}}{\lambda}} - t_e\right) dt_e$$

$$= \lambda_R \sqrt{\frac{2\varphi_{th}}{\lambda}} \int_{t_e=0}^{\sqrt{\frac{2\varphi_{th}}{\lambda}}} f_e(t_e) dt_e - \lambda_R \int_{t_e=0}^{\sqrt{\frac{2\varphi_{th}}{\lambda}}} f_e(t_e) t_e dt_e$$

$$= F_e\left(\sqrt{\frac{2\varphi_{th}}{\lambda}}\right) - F_e\left(\sqrt{\frac{2\varphi_{th}}{\lambda}}\right) + \lambda_R \int_{t_e=0}^{\sqrt{\frac{2\varphi_{th}}{\lambda}}} F_e(t_e) dt_e$$

$$= \sqrt{\frac{\lambda}{2\varphi_{th}}} \int_{t_e=0}^{\sqrt{\frac{2\varphi_{th}}{\lambda}}} F_e(t_e) dt_e$$

(1) becomes,

$$P_{DKGA} = 1 - \sqrt{\frac{\lambda}{2\varphi_{th}}} \int_{t_e=0}^{\sqrt{\frac{2\varphi_{th}}{\lambda}}} F_e(t_e) dt_e \quad (5)$$

This analytical model offers several benefits such as it is helpful in determining the key refresh-time for the specified scenarios, determining the type of key generation for specific requirements and the probability of dynamic key generation time under various constraints respectively. Also, it is recommended for the system to have a long key refresh time as the keys are very secure and can be used for a prolonged time to generate further keys using any key generation algorithms. By this model, it is possible to track the scenario where QKG-AKA can provide rigid security and adaptable to key generation. Through this analysis, we infer that deterministic algorithms cannot break the key hierarchy in polynomial time.

## VI. Performance Analysis

The proposed system was experimentally tested using five different servers running Ubuntu and windows. The experimental setups include the Quantum Key distributor as a program containing 12,987 unique key sets in a Core i3 machine and terminal nodes configured as UE in 3 different Core-i5 PC's running on a 2.50 GHz Intel ® Core™ i5-3120M Processor and 8 GB RAM. With this setup, the security and performance of the proposed mechanism is analyzed to estimate the computation and key generation time. The time taken for generating 256 bits through QKG (Quantum Key Generation) process is shown in Fig. 6. This proves the time taken for the normal key generation and QKG process are similar when they are used conversely with KDF & random function. This paves the way to adopt a QKG process to the places where a Complex KDF is used to generate a key. The authentication time taken for HSS and MME is analyzed and plotted in Fig. 7 and Fig. 8. The authentication time of various techniques such as QKG-AKA, SE-AKA[31] and EPS-AKA[32] and parameters have been compared with our mechanism. The X-axis represents the number of authentication and Y-axis signifies the elapsed time (seconds) of various mechanisms. The result clearly shows that the elapsed time of QKG-AKA mechanism is reduced when compared with other techniques without any security negotiations. From the QK-GRID, the number of unique keys generated is shown in the Fig. 9. We infer from the plot that the number of unique keys increases with the size of the QK-GRID.

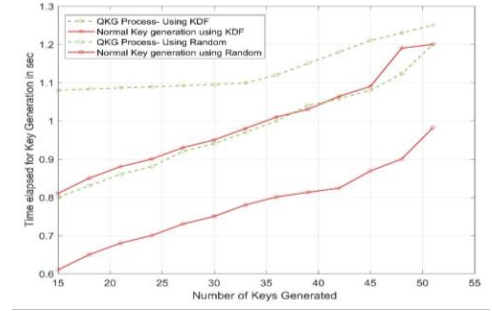

Fig. 6 Time taken for 256 bits Key generation by QKG process

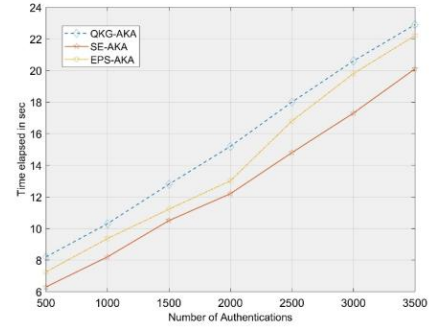

Fig. 7 Authentication load at HSS

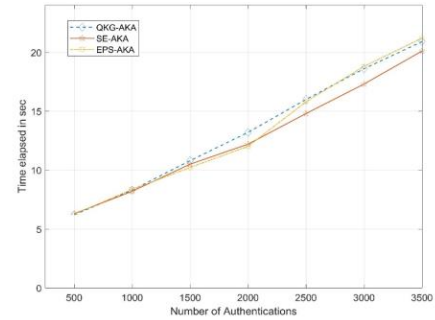

Fig. 8 Authentication load at MME

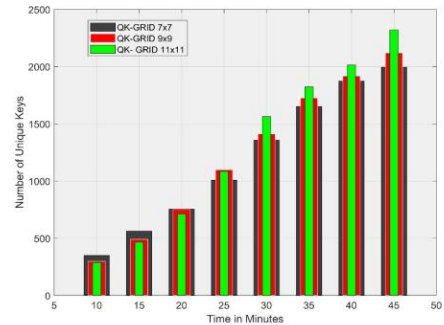

Fig. 9 Number of unique keys found for QK-GRID

## VII. Conclusion

The transmission medium is vulnerable to attacks due to its broadcasting nature and the features of the radio propagation channels. In the 5G era, it is expected that billions of devices are going be connected to the Internet and the backbone network is believed to be LTE-A network. Though technology development provides comfort, it also paves the ways for intruders. Attacks like jamming, privacy theft, eavesdropping etc., are threatening the evolving technologies. These attacks on the LTE-A network invites fruitful solutions to protect the network from adversaries. One of the predominant issues that seek immediate attention is the limitation in the authentication process and permanent security association. This has been resolved here by a secure framework, QKG-AKA using dynamic key hierarchy and quantum information. In this paper, we proposed QK-GRID based on QKD to overcome single static security association through a reliable key management process, with minimal architectural changes in LTE / SAE security architecture. The performance of the QKG-AKA has been tested with various parameters and found to be secure in all the attacker models. The analytical model of the proposed key hierarchy ensures that the probability of the hacking the system in polynomial time is not possible. We infer a twofold improvement in the security of the key hierarchy when it is deployed with QKD. The future work of this article is to extend the QK-GRID of the QKG-AKA to provide security for the Internet of Vehicles domain for a safe and smart transportation system.

ACKNOWLEDGMENT

This work was supported by the Deanship of Scientific Research (DSR), King Abdulaziz University, Jeddah, under grant No. (DF-241-611-1441). The authors, therefore, gratefully acknowledge DSR technical and financial support.

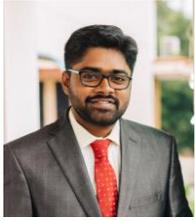

**RAJAKUMAR ARUL (M'15)** is currently working as Assistant Professor at the Department of Computer Science and Engineering, Amrita School of Engineering, Bengaluru. He pursued his Bachelor and Master's in Computer Science and Engineering from Anna University, Chennai. He completed his Doctorate of Philosophy requirements in Full Time under the Faculty of Information and Communication, Department of Computer Technology, Anna University - MIT Campus. He is a recipient of Anna Centenary Research Fellowship (ACRF) for his doctoral studies. His research interests include Security in Broadband Wireless Networks, Block Chain, WiMAX, LTE, Robust resource allocation schemes in Mobile Communication Networks. He is a member of IEEE and Professional member of ACM.

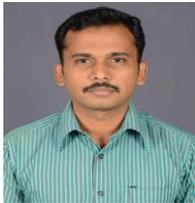

**GUNASEKARAN RAJA (M'08-SM'17)** received the Ph.D. degree from the Faculty of Information and Communication Engineering, Anna University, Chennai, India, in 2010. He is a Professor and Head of the Department of Computer Technology, Anna University, where he is also a Principal Investigator of NGNLab. He was a Postdoctoral Fellow with the University of California, Davis, CA, USA. His current research interests include 5G networks, Internet of Vehicles, Internet of Drones, wireless security, Machine Learning and data offloading. He was the recipient of the Young Engineer Award from the Institution of Engineers India in 2009, FastTrack grant for Young Scientist from the Department of Science and Technology in 2011, the Professional Achievement Award for the year 2017 from IEEE Madras Section, and Visvesvaraya Young Faculty Research Fellowship Award for the year 2019 from the Government of India.

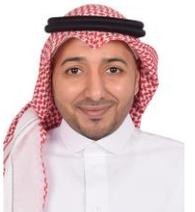

**ALAA OMRAN ALMAGRABI** received B.Sc (Computer science) degree from Jeddah teaching College in 2003 and the Master degree in Information Technology in 2009 and Ph.D (Computer Science in 2014) from La Trobe University in Melbourne, Australia. In 2014, he was appointed as Assistant Professor with the Department of Information System in the Computer Science and Information Technology at the University of King Abdulaziz in Jeddah, Saudi Arabia. In 2019, he was promoted to Associate Professor rank in the department of Information System. My research area includes Pervasive Computing, Networking, Data mining, System analysis and design, and ontology domains.

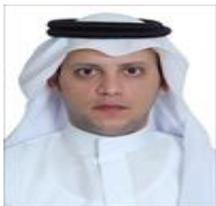

**MOHAMMED SAEED ALKATHEIRI** received the bachelor's degree in computer science from King Abdulaziz University, Jeddah, Saudi Arabia, the master's degree in communication network security from the Beijing University of Aeronau- tics and Astronautics (Beihang), China, and the Ph.D. degree in computer science from Texas Tech University, USA, through a full scholarship.
He was a Researcher with the Center of Excellence in Information Assurance, King Saud University, Riyadh, Saudi Arabia. He is currently an Assistant Professor with the Department of Cybersecurity, College of Computer Science and Engineering, University of Jeddah, Saudi Arabia, where he is also a Vice-Dean for Graduate Studies and Scientific Research. His current research interests focus on the area of cybersecurity, digital authentication, machine learning and pattern recognition, security in resource-constraint devices, and technological innovation management. He served as a Consultant for national projects and joined the Prince Muqrin Chair for Information Security Technology (PMC) along with government departments on National Information Security Strategy project as a Security Consultant.

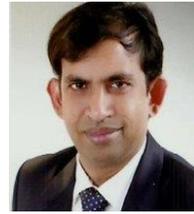

**CHAUHDARY SAJJAD HUSSAIN** is Assistant Professor at College of Computer Science and Engineering, University of Jeddah. He was Sr. Research Engineer at LG/LSIS Co., LTD. | Advance Technology R&D Center , joined in (2011), where he received "Best Research Award" (2012). He received his Ph.D. (2013) degree from Korea University , awarded Scholarship by Korea Government (2007) for Ph.D. studies.During his Ph.D. studies, he was Teaching Assistant and Research Assistant with Prof. Park Myong Soon (2007~2011). In (2006) he received M.S. degree from Ajou University, South Korea. During his Master studies, he was associated with Ubiquitous Korea Project, which was funded by Ministry of Knowledge Economy Korea, total amount of 13.345 billion KRW. In (2008), he joined SK Telecom as Visiting Researcher/Intern. He is member of major standardization organizations i.e. SAE International, ZigBee, ISO/IEC . His research interests include Cyber Security, Industrial Internet of Things, and Communication Infrastructure for Electric Vehicle and its Charging, Power Line Communications, Industrial Wired/Wireless Communication Networks.

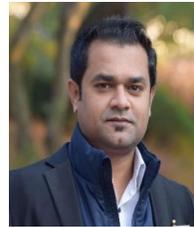

**ALI KASHIF BASHIR** is a Senior Lecturer at the Department of Computing and Mathematics, Manchester Metropolitan University, United Kingdom. He is a senior member of IEEE, invited member of IEEE Industrial Electronic Society, member of ACM, and Distinguished Speaker of ACM. His past assignments include Associate Professor of ICT, University of the Faroe Islands, Denmark; Osaka University, Japan (71 in QS Ranking 2020); Nara National College of Technology, Japan; the National Fusion Research Institute, South Korea; Southern Power Company Ltd., South Korea, and the Seoul Metropolitan Government, South Korea. He has worked on several research and industrial projects of South Korean, Japanese and European agencies and Government Ministries. He is also advising several start-ups in the field of STEM based education, block chain, robotics, and smart homes. He received his Ph.D. in computer science and engineering from Korea University (83 in QS Ranking 2020) South Korea. He is supervising/co-supervising several graduate (MS and PhD) students. His research interests include internet of things, wireless networks, distributed systems, network/cyber security, cloud/network function virtualization, etc. He is serving as the Editor-in-chief of the IEEE FUTURE DIRECTIONS NEWSLETTER.